\documentclass[lettersize, journal, twocolumn]{IEEEtran}
\usepackage{amsmath,amsfonts}
\usepackage{algorithmic}
\usepackage{algorithm}
\usepackage{array}
\usepackage[caption=false,font=normalsize,labelfont=sf,textfont=sf]{subfig}
\usepackage{textcomp}
\usepackage{stfloats}
\usepackage{url}
\usepackage{verbatim}
\usepackage{graphicx}
\usepackage{cite}
\usepackage{comment}
\usepackage{booktabs} 
\usepackage{makecell}
\usepackage{tabularx}
\usepackage[hidelinks]{hyperref}

\begin{document}
\title{Continuous Blood Monitoring with Particle-based Integrated Sensing and Communication (ISAC)}

\newcommand{\orcidiconFeb}{\href{https://orcid.org/0009-0008-2632-1140}{\includegraphics[scale=0.1]{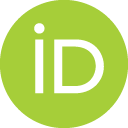}}}

\newcommand{\orcidiconOba}{\href{https://orcid.org/0000-0003-2523-3858}{\includegraphics[scale=0.1]{figures/orcidID128.png}}}

\author{Fatih E. Bilgen\orcidiconFeb,~\IEEEmembership{Graduate Student Member,~IEEE},
        and Ozgur B. Akan\orcidiconOba,~\IEEEmembership{Fellow,~IEEE}     
        \thanks{Fatih E. Bilgen, Ozgur B. Akan are with the Internet of Everything (IoE) Group, Electrical Engineering Division, Department of Engineering, University of Cambridge, Cambridge, CB3 0FA, UK (email: \{feb49, oba21\}@cam.ac.uk).}
        \thanks{Ozgur B. Akan is also with the Center for neXt-generation Communications (CXC), Department of Electrical and Electronics Engineering, Koç University, Istanbul, Turkey (e-mail: \{akan\}@ku.edu.tr).}
	    \thanks{This work was supported in part by the AXA Research Fund (AXA Chair for Internet of Everything at Ko\c{c} University).}
}

\maketitle
\thispagestyle{empty}
\begin{abstract} Although the circulatory system functions as a continuous source of physiological data, contemporary diagnostics remain bound to intermittent, time-delayed assessments. To resolve this, we present a framework for ubiquitous hematological profiling driven by Integrated Sensing and Communication (ISAC). We demonstrate how electromagnetic signals can be exploited to monitor blood in real-time, effectively converting them into diagnostic tools. We analyze the biological foundations of blood, review existing Complete Blood Count (CBC) and sensing technologies, and detail a novel pipeline for continuous blood monitoring. Furthermore, we discuss the potential applications of deploying these devices to enable real-time CBC and biomarker detection, ultimately revolutionizing how we predict, detect, and manage individual and public health. 
\end{abstract}


\section{Introduction}
\IEEEPARstart{T}he human circulatory system serves as a dynamic reservoir of physiological data, containing a complex matrix of cells, biomarkers, and potential pathogens. Despite this wealth of information, current medical practice lacks the infrastructure for continuous hematological monitoring in both clinical and everyday settings.

In inpatient environments, the absence of real-time blood analysis stands in stark contrast to the ubiquity of cardiac monitors. This disparity prevents clinicians from observing immediate fluctuations in a patient's condition, often delaying necessary intervention. Outside the hospital, reliance remains on a reactive snapshot model: a patient experiences symptoms, seeks care, and undergoes isolated testing. This sporadic approach fails to capture the temporal dynamics of disease progression from its asymptomatic onset. Consequently, metabolic disorders and pathogenic infections often develop unnoticed, bypassing the optimal window for early treatment.

To bridge this gap, there is a clear necessity for continuous blood monitoring systems tailored to diverse environments. In clinical settings, such devices could function analogously to cardiac monitors for hospitalized patients. For the general population, the focus shifts to wearable technologies, such as epidermal patches, designed to detect early warning signs during daily activities. Looking further ahead, we can envision a future where nanorobots equipped with nanoantennas circulate within the bloodstream, integrating sensing capabilities directly inside the body \cite{kuscu2021iobnt, akan2023IoE}.

Realizing this multi-scale vision requires a transformative technological framework. Integrated Sensing and Communication (ISAC) offers a pathway to achieve this by repurposing communication signals to monitor the blood medium across these varying scales. This article outlines a vision for continuous, always-on hematology, by exploring how ISAC techniques can create a digital picture of blood.

The remainder of this paper is organized as follows. Section \ref{sec:2} examines the composition of human blood tissue and identifies its fundamental constituents. Section \ref{sec:3} provides a review of state-of-the-art technologies in hematology, specifically addressing recent developments in CBC and biomarker-based detection. Section \ref{sec:4} introduces our novel framework and pipeline for continuous blood monitoring. Finally, Section \ref{sec:5} outlines potential applications of this technology, followed by concluding remarks in Section \ref{sec:6}.

\section{The Composition of Human Blood Tissue}
\label{sec:2}
Human blood is a unique type of connective tissue that acts as the physical medium for carrying gases, nutrients, hormones, and waste products throughout the body. It consists of two primary components: plasma, the liquid matrix forming about 55\% of the blood's volume, and the formed elements, which make up the remaining 45\% and represent the cellular part of blood. These components together sustain physiological balance and offer a dynamic insight into health \cite{hall2015guyton}.

\subsection{The Formed Elements of Blood}
\begin{figure*}[!htbp]
    \centering
    \includegraphics[width=1\linewidth]{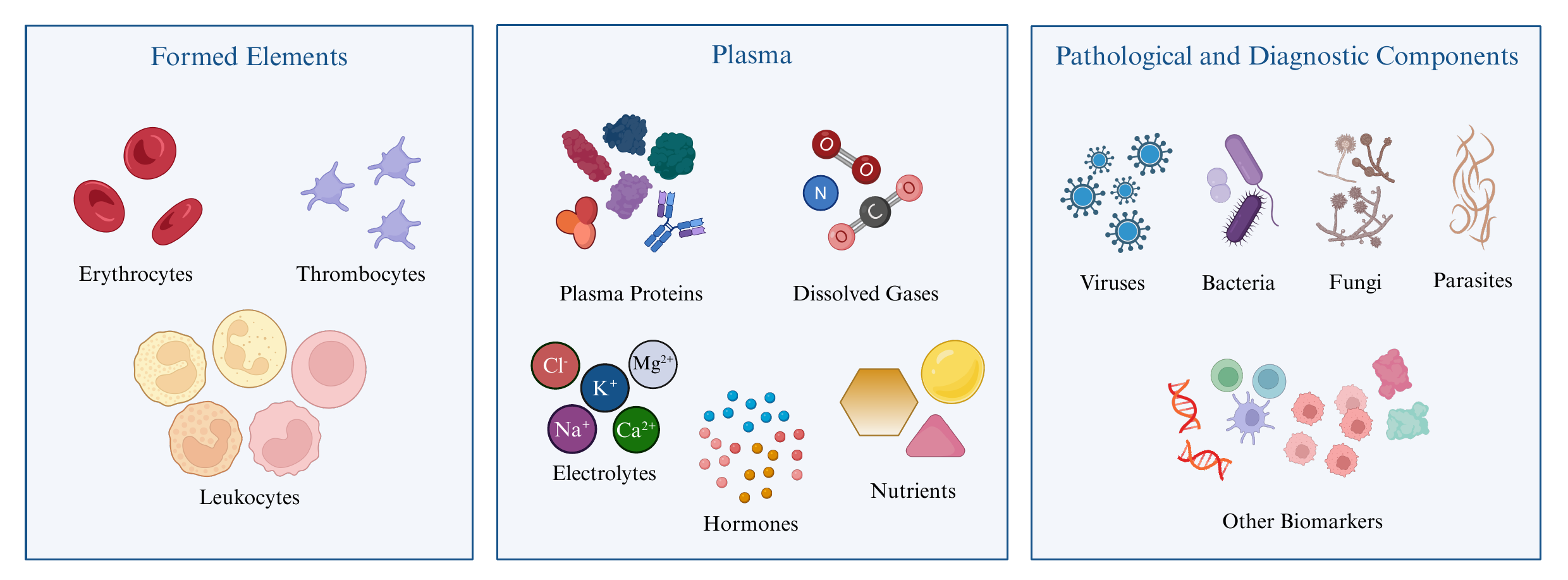}
    \caption{Schematic overview of human blood constituents. The figure classifies blood components into three distinct categories: \textbf{Formed Elements}, comprising the cellular fraction (erythrocytes, thrombocytes, and leukocytes); \textbf{Plasma}, the liquid matrix containing proteins, dissolved gases, electrolytes, hormones, and nutrients; and \textbf{Pathological and Diagnostic Components}, which illustrates foreign pathogens (viruses, bacteria, fungi, parasites) and various biomarkers utilized for clinical health assessment \cite{Bilgen2026Fig1}.}
    \label{fig:1}
\end{figure*}

The formed elements of blood consist of erythrocytes, leukocytes, and thrombocytes (Fig. \ref{fig:1}), each carrying out unique physiological functions \cite{kvrivzkova2021blood}.

Erythrocytes, or red blood cells, account for about 99\% of all blood cells and are primarily responsible for transporting oxygen and carbon dioxide via hemoglobin. These mature cells are biconcave, lack a nucleus, and possess a lifespan of approximately 120 days, features that facilitate optimal gas exchange and deformability within capillaries \cite{kvrivzkova2021blood}. 

Leukocytes, or white blood cells, constitute less than 1\% of total blood cells yet are critical for immune defense and tissue repair. These cells are classified by cytoplasmic granularity into two main groups: granulocytes and agranulocytes \cite{kvrivzkova2021blood}. Granulocytes include neutrophils (60--70\% of leukocytes), which phagocytose bacteria and fungi and possess a multilobed nucleus; eosinophils (2--4\%), identifiable by their bilobed nuclei and orange-red granules, which defend against parasites and modulate allergic reactions; and basophils ($<$1\%), featuring large blue-purple granules that often obscure the nucleus, responsible for releasing histamine and heparin during inflammation and allergic responses. 

Complementing the granulocytes are the agranulocytes, which include lymphocytes and monocytes. Lymphocytes (20--30\%) orchestrate adaptive and innate immune responses through B cells for antibody production, T cells for immune regulation and cytotoxicity, and Natural Killer (NK) cells for nonspecific cytotoxic activity. Monocytes (2--8\%) are the largest blood cells and differentiate into macrophages and dendritic cells in tissues, enabling phagocytosis and antigen presentation.

Finally, hemostasis is maintained by thrombocytes, or platelets. These small, anucleate cytoplasmic fragments are derived from megakaryocytes in the bone marrow and have an average lifespan of 7--10 days. Their main role is to initiate clot formation for wound repair \cite{kvrivzkova2021blood}. 

\subsection{Plasma: The Liquid Matrix}
Plasma, the straw-colored liquid portion of blood, comprises approximately 90\% water and serves as the solvent suspending the formed elements (Fig. \ref{fig:1}). It contains a diverse array of dissolved substances, including proteins, solutes, and small molecules that are vital for physiological homeostasis \cite{hall2015guyton}.

Plasma proteins, constituting 7--8\% of the plasma, perform essential roles. Albumin, which makes up roughly 60\% of these proteins, maintains osmotic pressure and facilitates the transport of lipids and hormones. Globulins ($\sim$35\%), comprising $\alpha$-, $\beta$-, and $\gamma$-globulins, mediate immune defense and transport metals and lipids. Fibrinogen ($\sim$4\%) acts as the precursor to fibrin in the coagulation pathway, while regulatory proteins ($\sim$1\%) encompass enzymes, proenzymes, and complement proteins involved in catalytic and immune processes. Additionally, antibodies or immunoglobulins (IgG, IgA, IgM, IgE, and IgD) secreted by plasma cells bind antigens to mediate immune responses \cite{hall2015guyton}.

Beyond proteins, plasma serves as a complex transport medium. It carries nutrients absorbed from digestion, such as carbohydrates, lipids, amino acids, and vitamins, alongside electrolytes like Na$^+$, K$^+$, Ca$^{2+}$, Mg$^{2+}$, Cl$^-$, HCO$_3^-$, and PO$_4^{3-}$, which regulate osmotic balance, pH, membrane potential, and muscle contraction. Hormones produced by endocrine glands, including peptide, steroid, and amine hormones, are transported in plasma to regulate systemic physiology. Simultaneously, dissolved gases like oxygen, carbon dioxide, and nitrogen are conveyed in various chemical forms for respiration and metabolism. The plasma also facilitates the removal of metabolic wastes, carrying urea, uric acid, creatinine, bilirubin, and lactic acid to excretory organs. Additional molecules present include buffer systems, transport proteins, complement components, and clotting factors \cite{hall2015guyton}. 

\subsection{Pathological and Diagnostic Components}
In addition to its physiological components, blood can include foreign or abnormal elements that act as diagnostic markers for diseases (Fig. \ref{fig:1}). Blood acts as a reservoir for foreign intruders, harboring viruses, bacteria, parasites, fungi, and rarely, prions \cite{foucar2023diagnostic}. 

Beyond pathogens, specific biomarkers provide critical data on internal pathologies. Neoplastic markers such as CEA, AFP, CA-125, PSA, and LDH serve in the early detection and monitoring of cancers. Metabolic and organ function are evaluated through parameters like glucose, HbA1c, cholesterol, triglycerides, liver enzymes, creatinine, BUN, and uric acid, while endocrine disorders are identified via hormones like cortisol and ACTH. Furthermore, indicators such as CRP, ESR, ANA, rheumatoid factor, and cytokines reflect immune dysregulation and inflammation \cite{blann2022blood}.

Advanced analysis also reveals genetic, hematologic, and toxicological states. Circulating cell-free DNA, oncogene mutations, chromosomal abnormalities, and microRNAs provide molecular insights into hereditary and neoplastic diseases. Routine hematologic parameters, including hemoglobin, hematocrit, and coagulation factors, inform anemia classification and clotting status. Finally, blood analysis detects exogenous substances, ranging from therapeutic drugs and lifestyle-related metabolites to toxins like lead and mercury \cite{blann2022blood}.

\section{State-of-the-Art in Blood Monitoring Technologies}
\label{sec:3}

The contemporary convergence of decentralized immediacy and centralized efficiency defines the worldwide diagnostic landscape. The haematology market, valued at USD 4.33 billion in 2025 and projected to reach USD 7.28 billion by 2034, is a good indicator of general diagnostic trends \cite{ozelleseo_2025_top}. A sector that was previously rigidly separated into basic Point-of-Care (PoC) screening and high-throughput central laboratories is now becoming a hybrid environment. This shift is powered by the intersection of three complementary technological pillars: microfluidics, artificial intelligence (AI), and next-generation biosensing materials. Together, these innovations are driving a transition away from centralized, infrequent testing toward decentralized and minimally invasive monitoring \cite{__hematology}.

This section analyzes the current state-of-the-art technologies that characterize this field. It assesses the sustained dominance and ongoing refinement of automated hematology analyzers developed by major industry players, comparing their established, high-level performance with the disruptive potential of new PoC platforms. 

\subsection{Advanced Automated Laboratory Hematology}

As decentralized testing continues to grow, high-throughput automated hematology analyzers still serve as the clinical gold standard. At present, this sector is defined by a rivalry between two fundamentally different technological philosophies regarding cellular characterization: Fluorescence Flow Cytometry (FFC), pioneered by Sysmex Corporation, and morphometric Volume-Conductivity-Scatter (VCS), pioneered by Beckman Coulter \cite{serrandoquerol_2021_evaluation}.

The fluorescence-based strategy, utilized in the Sysmex XN-Series, evolves from conventional impedance counting by utilizing specific lysing reagents to perforate cell membranes and proprietary fluorescent dyes to stain intracellular nucleic acids. In contrast, Beckman Coulter’s VCS technique avoids harsh cytoplasmic stripping to assess cells in their near-native state. This approach combines electrical impedance to measure volume, radiofrequency signals to evaluate internal conductivity, and multi-angle light scatter to probe surface morphology \cite{yu_2023_performance}.

These competing architectures dictate the analytical strengths of each platform. Sysmex’s fluorescence systems rely on signal intensity proportional to nucleic acid content, a method particularly advantageous for identifying immature cells with high RNA levels and quantifying reticulocytes (immature RBCs) \cite{serrandoquerol_2021_evaluation}. Furthermore, a dedicated channel for platelets utilizes dyes specific to the endoplasmic reticulum, allowing for accurate counts even in thrombocytopenic samples where electrical impedance might fail due to interference. Conversely, the morphometric approach employed by Beckman Coulter emphasizes the preservation of cellular morphology; conductivity measurements distinguish cells of similar size but different internal composition, while scatter analysis determines granularity without the need for complex staining \cite{yu_2023_performance}.

Table \ref{tab:auto_hem_tech} highlights critical operational distinctions between the most recent technology of these flagship platforms. In this table, Maximum System Throughput/Scalability refer to an analyzer's capacity to handle increasing workloads by connecting multiple units together. High-volume mega-labs processing thousands of samples daily require systems that can scale without creating bottlenecks. In this regard, the Sysmex XN-9100 is designed as a modular track system capable of linking over nine analysis modules to achieve a massive throughput exceeding 900 samples per hour \cite{__xn9100}. Conversely, the Beckman Coulter DxH 900 is typically deployed in smaller workcells of up to three connected units, capping at approximately 300 samples per hour \cite{__dxh}. Sample Volume represents the minimum amount of whole blood required to perform a complete analysis, a crucial metric for pediatric and oncology care where patients often have difficult venous access or are prone to anemia from frequent blood draws. The Sysmex XN-9100 is significantly more efficient in this domain, requiring only 88 $\mu$L of whole blood. The DxH 900 requires nearly double that volume (165 $\mu$L).

\begin{table}[!t]
\centering
\caption{Comparison of Automated Hematology Analyzers}
\label{tab:auto_hem_tech}
\begin{tabular}{@{}lll@{}}
\toprule
\textbf{Feature} & \makecell[l]{\textbf{Sysmex} \\ \textbf{XN-9100}} & \makecell[l]{\textbf{Beckman Coulter} \\ \textbf{DxH 900}} \\ \midrule
\textbf{Max Throughput} & $>$900 samples/hr & 300 samples/hr \\ \midrule
\textbf{Sample Volume} & 88 $\mu$L & 165 $\mu$L \\ \midrule
\textbf{Technology} & FFC & VCSn \\ \midrule
\end{tabular}
\end{table}

\subsection{Point-of-Care (PoC) for CBC}
While central laboratories focus on throughput and parameter depth, a parallel revolution is occurring at the Point-of-Care (PoC), where the historical trade-off between portability and analytical performance is being eliminated. A new class of micro-hematology platforms has emerged that rivals reference analyzers (Table \ref{tab:poc_cbc}), specifically by delivering a 5-part differential, a breakdown of white blood cells into their five major types (neutrophils, lymphocytes, monocytes, eosinophils, and basophils) to provide a comprehensive immune profile.

Leading this charge is the PixCell HemoScreen, which achieves reference-grade accuracy through the application of microfluidic Viscoelastic Focusing (VEF) \cite{bransky_2021_novel}. Unlike traditional systems that rely on bulky pumps and complex sheath-flow fluids, this device mixes the blood sample with a proprietary viscoelastic buffer. The interaction between the fluid’s elastic properties and the cells generates lift forces that drive the cells toward the microchannel’s centerline, forming a single-file monolayer for interrogation. This efficient, self-aligning mechanism allows the HemoScreen to process a single drop of blood ($\sim$20 $\mu L$) and deliver 22 parameters in approximately five minutes \cite{__hemoscreen}.

Taking a different approach, the Sight OLO eliminates flow entirely. Instead of moving cells past a sensor, it utilizes Digital Blood Counting to digitize a static blood monolayer derived from roughly two drops of blood ($\sim$27 $\mu L$). The user creates a smear within a disposable cartridge, which the device then scans using multispectral imaging. This method allows AI algorithms to identify and count individual cells based on their visual signature rather than laser scatter, effectively treating blood analysis as a computer vision problem. This digital methodology yields 19 parameters with a 5-part differential in about 10 minutes \cite{bachar_2021_artificial}.

Distinct from the others in its focus on morphology is the Noul miLab, which features a solid-state staining method termed NGSI. This platform automates microscopy by using a hydrogel stamp to stain cells from a minimal sample volume of $\sim$4 $\mu L$. While the analysis time is longer, ranging from 15 to 20 minutes, the miLab offers specialized diagnostic value by identifying malaria species alongside standard CBC parameters \cite{_2023_blood}.

\begin{table*}[ht]
\centering
\caption{Comparative Analysis of Point-of-Care CBC Devices}
\label{tab:poc_cbc}
\renewcommand{\arraystretch}{1.3} 
\begin{tabular}{@{}llll@{}}
\toprule
\textbf{Feature} & \textbf{Sight OLO} & \textbf{PixCell HemoScreen} & \textbf{Noul miLab BCM} \\ \midrule
\textbf{Methodology} & Digital Imaging + AI & Viscoelastic Focusing (VEF) & NGSI Solid-State Staining \\ \midrule
\textbf{Sample Vol.} & $\sim$27 $\mu$L & $\sim$20 $\mu$L & $\sim$4 $\mu$L \\ \midrule
\textbf{Time} & $\sim$10 mins & $\sim$5 mins & $\sim$15-20 mins \\ \midrule
\end{tabular}
\end{table*}

\begin{table*}[!b]
\centering
\caption{Economic and Clinical Comparison: Central Lab vs. PoC}
\label{tab:lab_vs_poc}
\renewcommand{\arraystretch}{1.4}
\begin{tabularx}{\textwidth}{@{}lXX@{}}
\toprule
\textbf{Feature} & \textbf{Central Laboratory Analyzers} & \textbf{Point-of-Care (PoC) Devices} \\ \midrule
\textbf{Capital Model} & High (\$80k--150k). & Low (\$15k--30k). Low barrier to entry. \\ \midrule
\textbf{Operational Cost} & Low CPT (\$1.00--3.00). Cost driven by labor \& service. & High CPT (\$10--30). Cost driven by single-use cartridges. \\ \midrule
\textbf{RoI Driver} & Volume efficiency and soft savings (logistics/safety). & Clinical pathway efficiency. \\ \midrule
\textbf{Sample Type} & Venous blood. & Capillary (fingerprick) validated as comparable to venous. \\ \midrule
\textbf{Validation} & Reference Standard for all parameters. & High correlation for core parameters. \\ \midrule
\textbf{Limitations} & Turnaround time and logistical delays. & Specific biases for some parameters. \\ \bottomrule
\end{tabularx}
\end{table*}

\subsection{Laboratory Hematology vs. Point-of-Care}
Moving testing from a central lab to the patient's bedside assumes that the results will be just as accurate, but the reality is slightly more complex. Generally, leading PoC devices perform very well; research shows they achieve high correlation ($r > 0.95$) for standard counts like white blood cells, red blood cells, and platelets when compared to large lab machines \cite{bachar_2021_artificial}. 

They also achieve high standards for ruling out conditions using biomarkers \cite{ellis_2024_performance}. Another major advantage confirmed by these studies is that capillary blood taken from a simple fingerprick provides results that are statistically comparable to a traditional venous blood draw. 

However, these devices are not perfect replacements in every scenario. While some devices offer excellent precision, some devices report lower monocyte counts or show slight dips in Hematocrit and MCV levels compared to reference methods \cite{shean__clinical}.

Financially, central laboratories and PoC devices operate on divergent economic models driven by total cost of ownership rather than simple capital expenditure, as summarized in Table \ref{tab:lab_vs_poc}. Central laboratories are built on high-volume automation and premium AI-integrated platforms. The initial capital investment for these advanced systems is substantial, typically ranging between USD 80,000 and over USD 150,000 \cite{ozelleseo_2026_cell}. However, this heavy upfront cost is offset by superior economies of scale. In high-volume settings processing over 125,000 tests annually, the comprehensive Cost Per Test (CPT) drops dramatically to approximately USD 1.00 to USD 3.00. In this setting, value is derived not just from consumable efficiency, but from hidden operational savings. Modern AI algorithms reduce the need for manual microscopic review, generating significant labor savings that are unavailable in legacy impedance-based systems \cite{ozelleseo_2026_cell}.

In contrast, PoC and entry-level devices utilize a financial model focused on accessibility and rapid deployment. These units are relatively affordable to acquire, with capital costs in 2025 generally falling between USD 15,000 and USD 30,000 \cite{ozelleseo_2026_cell}. However, the operational economics differ sharply from central labs due to lower throughput and reliance on single-use cartridge technology. Consequently, the daily running costs are significantly higher, with the total CPT hovering between USD 10.00 and USD 30.00. Despite this higher operational expense, these devices remain financially viable for smaller clinics processing 20 to 50 samples daily.

Because the per-test cost is so much higher for PoC, the Return on Investment (RoI) does not come from saving money on the test itself, but from saving time in the clinical process. The value becomes clear when a fast result solves a logistical problem. While the central lab is still the most cost-effective choice for routine testing, the premium price of PoC is easily justified whenever it speeds up patient flow, halts rapid disease progression, or prevents operational bottlenecks that are far more expensive than a disposable cartridge.

\section{ISAC for Continuous Blood Monitoring}
\label{sec:4}

\begin{figure*}[!htbp]
    \centering
    \includegraphics[width=1\linewidth]{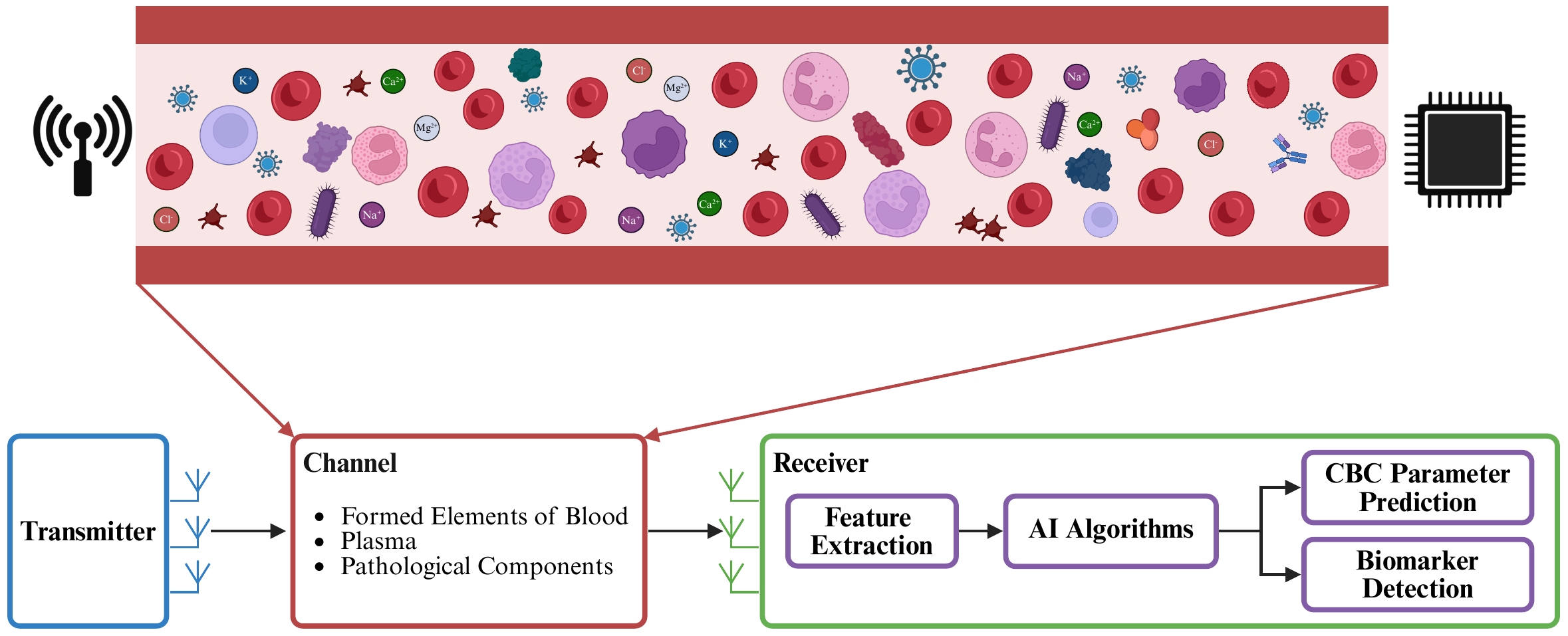}
    \caption{Continuous Blood Monitoring with Integrated Sensing and Communication system architecture showing signal processing pipeline \cite{Bilgen2026Fig2}.}
    \label{fig:2}
\end{figure*}

Despite the transformative capabilities of the PoC platforms discussed previously, a critical gap remains: the lack of temporal resolution. The analysis of blood components remains a cornerstone of medical diagnostics, where variations in cell counts and the presence of abnormal cells provide the primary signals for detecting functional disorders and specific diseases. However, even the most sensitive techniques or the most precise micro-hematology analyzers are limited to discrete, isolated measurements. Currently, no methodology exists to monitor blood composition continuously, leaving a blind spot for transient physiological events that occur between screenings.

To bridge the gap between discrete diagnostics and continuous monitoring, we propose utilizing Integrated Sensing and Communication (ISAC). This approach is grounded in the principle that blood constituents and pathogens possess distinct dielectric, physical, and chemical profiles. Consequently, distinct cell types interact uniquely with propagating communication signals, embedding identifiable fingerprints into the transmission channel. By extracting and decoding these channel variations, receivers can derive real-time insights into the continuous state of the biological medium.

Realizing this technology requires a systematic, multi-stage framework. First, a comprehensive library characterizing the properties of individual biological cells need to be established to serve as the ground truth for sensing. Using this data, a multiphysics simulation environment can be constructed to model the complex interactions between these cells and communication signals under controlled conditions. Through these simulations, channel data can be extracted to isolate signal deviations correlated with specific disorders. Finally, Artificial Intelligence (AI) models can be employed to decipher these complex patterns within the data, enabling the system to autonomously assess the state of the blood and trigger warnings in the event of physiological anomalies. Fig. \ref{fig:2} illustrates the proposed system architecture.

While our investigation focuses specifically on electromagnetic communication, the proposed framework is essentially independent of modality. We assume the use of electromagnetic transmitters and receivers for the current scope. However, future developments may extend this methodology to acoustic, molecular, or other communication modalities. Furthermore, the adoption of multi-modal transceivers would allow for simultaneous operation across domains, thereby increasing the number of extractable features and improving system accuracy.

\subsection{Electromagnetic Interaction Principles}
The feasibility of Integrated Sensing and Communication (ISAC) within the bloodstream relies on the distinct physical interactions between electromagnetic signals and biological entities \cite{liu2020ISAC}. Understanding these interactions is essential, as they define how cellular properties modify the electromagnetic signal, thereby serving as the basis for continuous sensing.

The interaction of electromagnetic waves with biological materials is fundamentally governed by three frequency-dependent properties: complex permittivity ($\epsilon$), complex permeability ($\mu$), and conductivity ($\sigma$) \cite{pozar_2012_microwave}. Permittivity defines the material's ability to store and dissipate electrical energy, while permeability relates to magnetic energy storage and loss. Conductivity quantifies the flow of electric charges and serves as the dominant absorption mechanism in conductive media.

From these fundamental parameters, critical derived metrics are established, including wave impedance, which dictates the partition of energy into reflected and transmitted components at boundaries, and skin depth, which characterizes the penetration distance before significant attenuation occurs \cite{pozar_2012_microwave}.

When incident waves encounter biological interfaces, the mismatch in wave impedance results in reflection, transmission, and refraction phenomena described by Snell's Law \cite{pozar_2012_microwave}. Crucially, these interactions exhibit strong frequency dependence, or dispersion.  The biological dielectric spectrum is segmented into distinct regions: the low-frequency $\alpha$-dispersion associated with counter-ion diffusion, the radio-frequency $\beta$-dispersion arising from interfacial polarization at the cell membrane, and the high-frequency $\gamma$-dispersion caused by the relaxation of water molecules. The $\beta$-dispersion is of particular diagnostic value; it originates from the Maxwell-Wagner effect, where charge carriers accumulate at the insulating membrane interface, effectively creating a large induced dipole that characterizes the cell's structural integrity \cite{martinsen2023bioimpedance}.

\subsection{Electromagnetic Characterization of Biological Cells}
To utilize the interaction principles described above for sensing, the spectral signatures of the biological channel must be quantitatively mapped to specific physical cell properties. This requires the definition of a biophysical dictionary that converts experimental electromagnetic data into intrinsic cellular parameters, which serve as the necessary inputs for the mathematical modeling of the channel.

Standard biophysical frameworks, such as the single-shell model, are employed for this purpose. This model conceptualizes the cell as a conductive cytoplasm enclosed by a thin, insulating membrane, allowing for the derivation of specific intrinsic values such as membrane capacitance ($C_m$) and cytoplasm conductivity ($\sigma_{cyto}$) \cite{martinsen2023bioimpedance}. Notably, micro-scale characterization methods focus exclusively on these dielectric properties. Because biological materials are overwhelmingly non-magnetic, techniques like Dielectrophoresis (DEP) and Impedance Cytometry rely entirely on the dielectric contrast between the cell and the suspending medium, rather than magnetic susceptibility.

However, a critical gap currently persists in the literature regarding the standardization of these parameters. The construction of a comprehensive electromagnetic library characterizing the full spectrum of blood constituents, spanning from healthy erythrocytes and leukocytes to rare pathological variants, remains an open research need. Establishing such a ground truth is a prerequisite for achieving high-fidelity simulations. Consequently, in the following section, we assume the existence of these characterized parameters to construct the mathematical framework for the communication channel.

\subsection{Mathematical Model}
We consider a cylindrical nano-communication channel of length L and radius R, filled with a biological medium consisting of a host fluid and a high concentration of suspended biological cells. The communication system is defined by a pair of nano-antennas located at opposite ends of the cylinder: a transmitter (Tx) that releases an electromagnetic pulse, and a receiver (Rx) that collects the propagated signal. The medium acts as a dense colloidal suspension where each cell is parameterized by its volumetric concentration, denoted by $\phi$, and its complex dielectric properties, $\epsilon_{cell}$. The interaction between the electromagnetic signal and this dense cellular environment forms the basis of the channel response.

To model the signal propagation, we adopt the Effective Medium Theory (EMT) approach \cite{choy_2015_effective}. EMT provides an analytical framework for determining a material’s macroscopic complex dielectric properties based on the characteristics and volume fractions of its individual components \cite{hernandez-cardoso_2020_empirical}. Rather than treating the heterogeneous mixture of cells and plasma as a collection of discrete scatterers, this method homogenizes the environment into a single, uniform material characterized by an effective complex permittivity, $\epsilon_{eff}$. This simplification allows us to derive the channel impulse response analytically while retaining the physical influence of the cellular parameters.

To characterize the channel impulse response, we must first derive the frequency-dependent complex permittivity of the biological medium. This is achieved in a two-step process: first, by modeling the dielectric response of individual cells, and second, by homogenizing the cellular suspension into an effective medium.

\subsubsection{Dielectric Modeling of Single Cells}
Biological cells are heterogeneous structures composed of conductive and dielectric layers. We utilize the multi-shell dielectric model to calculate the equivalent complex permittivity of a single cell, denoted as $\epsilon_{cell}^*(f)$.

The complex permittivity $\epsilon^*$ for any constituent material (cytoplasm, membrane, or plasma) is defined as
\begin{equation}
    \epsilon^*(f) = \epsilon_r'(f) - j \frac{\sigma(f)}{f \epsilon_0}, 
\end{equation}
where  $f$ is the linear frequency, $\epsilon_r'(f)$ is the frequency-dependent relative permittivity, $\sigma(f)$ is the frequency-dependent electrical conductivity, $\epsilon_0$ is the vacuum permittivity \cite{_2017_clausius_ch6}.

For non-nucleated cells, we employ a single-shell model consisting of a cytoplasm core surrounded by a thin plasma membrane \cite{_2017_clausius_ch9}. The equivalent permittivity of the cell, $\epsilon_{cell}^*$, is derived as
\begin{equation}
    \label{eq:single_cell_permittivity}
    \epsilon_{cell}^* = \epsilon_{mem}^* \frac{2\epsilon_{mem}^* + \epsilon_{cyto}^* + 2\nu(\epsilon_{cyto}^* - \epsilon_{mem}^*)}{2\epsilon_{mem}^* + \epsilon_{cyto}^* - \nu(\epsilon_{cyto}^* - \epsilon_{mem}^*)},
\end{equation}
where $\epsilon_{cyto}^*$ and $\epsilon_{mem}^*$ are the complex permittivities of the cytoplasm and membrane, $\nu = (1 - d/R)^3$ is the volume fraction of the core, with $R$ being the cell radius and $d$ the membrane thickness.

For nucleated cells, a double-shell model is necessary to account for the nucleus \cite{_2017_clausius_ch9}. This is calculated by applying Eq. (\ref{eq:single_cell_permittivity}) iteratively. First, we calculate the equivalent permittivity of the nucleated core ($\epsilon_{nuc\_eff}^*$), treating the nucleus as the core and the nuclear envelope as the shell
\begin{equation}
    \label{eq:double_shell_inner_permittivity}
    \epsilon_{nuc\_eff}^* = \epsilon_{env}^* \frac{2\epsilon_{env}^* + \epsilon_{nuc}^* + 2\nu_{nuc}(\epsilon_{nuc}^* - \epsilon_{env}^*)}{2\epsilon_{env}^* + \epsilon_{nuc}^* - \nu_{nuc}(\epsilon_{nuc}^* - \epsilon_{env}^*)},
\end{equation}
where $\epsilon_{nuc}^*$ and $\epsilon_{env}^*$ are the complex permittivities of the nucleus and nuclear envelope, and $\nu_{nuc} = (1 - d_{env}/R_{nuc})^3$ with $R_{nuc}$ being the nucleus radius and $d_{env}$ the nuclear envelope thickness. 

We then treat this nucleated core as the interior of the cytoplasm, surrounded by the outer membrane. Finally, the total cell permittivity $\epsilon_{cell}^*$ is found by substituting $\epsilon_{nuc\_eff}^*$ into Eq. (\ref{eq:single_cell_permittivity}) in place of $\epsilon_{cyto}^*$ as
\begin{equation}
    \label{eq:double_shell_cell_permittivity}
    \epsilon_{cell}^* = \epsilon_{mem}^* \frac{2\epsilon_{mem}^* + \epsilon_{nuc\_eff}^* + 2\nu(\epsilon_{nuc\_eff}^* - \epsilon_{mem}^*)}{2\epsilon_{mem}^* + \epsilon_{nuc\_eff}^* - \nu(\epsilon_{nuc\_eff}^* - \epsilon_{mem}^*)},
\end{equation}
where $\epsilon_{mem}^*$ is the complex permittivity of the outer membrane, $\epsilon_{nuc\_eff}^*$ defined in Eq. (\ref{eq:double_shell_inner_permittivity}),  $\nu= (1 - d/R)^3$ is the volume fraction of the core, with $R$ being the cell radius and $d$ the membrane thickness.

\subsubsection{Multicomponent Effective Medium Homogenization}
To accurately capture the heterogeneity of the biological channel, we model the blood as a complex colloidal suspension. We consider water as the continuous host medium ($\epsilon_{water}^*$) containing three distinct categories of inclusions: Formed Elements, consisting of $M$ distinct types indexed by $i$; Plasma Constituents (excluding water), consisting of $N$ distinct types indexed by $j$; and Pathological Components, consisting of $P$ distinct types indexed by $k$.

We employ the Generalized Maxwell-Garnett mixing rule \cite{choy_2015_effective}. To simplify the expression, we first define $\beta$, for any constituent $x$ relative to the water host as

\begin{equation}
    \label{eq:beta}
    \beta_x = \frac{\epsilon_{x}^* - \epsilon_{water}^*}{\epsilon_{x}^* + 2\epsilon_{water}^*}.
\end{equation}

The effective complex permittivity of the channel, $\epsilon_{eff}^*$, is then derived by summing the volume-weighted polarizabilities of all three categories as
\begin{equation}
    \label{eq:effective_medium}
    \small \epsilon_{eff}^* = \epsilon_{water}^* \frac{1 + 2 \left( \sum_{i=1}^{M} \phi_i \beta_i + \sum_{j=1}^{N} \phi_j \beta_j + \sum_{k=1}^{P} \phi_k \beta_k \right)}{1 - \left( \sum_{i=1}^{M} \phi_i \beta_i + \sum_{j=1}^{N} \phi_j \beta_j + \sum_{k=1}^{P} \phi_k \beta_k \right)}, 
\end{equation}
where $\phi_i, \phi_j$, and $\phi_k$ represent the volume fractions of the specific formed, plasma, and pathological elements, respectively, and $\beta_i, \beta_j, \beta_k$ are the calculated polarizability factors for each respective type using Eq. (\ref{eq:beta}).

\subsubsection{Channel Transfer Function and Impulse Response}
Having derived the effective macroscopic permittivity of the biological medium, we now formulate the signal propagation model. The cylindrical channel is treated as a lossy transmission medium where the signal degradation is governed by two primary mechanisms: geometric spreading of the electromagnetic wave and molecular absorption due to the effective dielectric properties of the homogenized mixture.

The interaction between the electromagnetic wave and the homogenized biological medium is characterized by the complex propagation constant, $\gamma(f)$. For a non-magnetic medium ($\mu_r = 1$), $\gamma(f)$ is directly derived from the effective complex permittivity $\epsilon_{eff}^*(f)$ calculated in Eq. (\ref{eq:effective_medium}) as
\begin{equation}
    \label{eq:propagation}
    \gamma(f) =  \alpha(f) + j\beta(f) = j\frac{f}{c} \sqrt{\epsilon_{eff}^*(f)} ,
\end{equation}
where we obtain the attenuation constant $\alpha(f)$ and the phase constant $\beta(f)$ \cite{pozar_2012_microwave}. 

The channel transfer function $H(f, L)$ describes how the amplitude and phase of the transmitted signal $X(f)$ are modified as they traverse the channel length $L$. The received signal spectrum is given by $Y(f) = H(f, L) X(f)$. We model $H(f, L)$ as the product of two distinct components as
\begin{equation}
    \label{eq:freq_response}
    H(f, L) = \left(\frac{c}{4\pi f L}\right)^{\eta/2} \cdot e^{-\gamma(f) L},
\end{equation}
where the first term represents the signal attenuation due to wavefront expansion and the second term accounts for the interaction with the biological medium \cite{pozar_2012_microwave}. Here, the exponent $\eta$ is the path loss exponent, where $\eta=2$ corresponds to free-space spherical spreading, and $\eta < 2$ indicates a guiding effect within the cylindrical vessel, and $\gamma(f)$ is given in Eq. (\ref{eq:propagation}), introducing frequency-selective attenuation via the real part and a phase delay via the imaginary part.

The time-domain behavior of the system can also be described by the Channel Impulse Response (CIR), $h(t)$. This is obtained by performing the Inverse Fourier Transform (IFT) of the transfer function
\begin{equation}
    h(t) = \mathcal{F}^{-1} \{ H(f, L) \} = \int_{-\infty}^{\infty} H(f, L) e^{j2\pi ft} df.
\end{equation}

\subsection{Simulation Framework and Feature Extraction}
The translation of theoretical cellular characteristics into detectable signals commences with the construction of a high-fidelity digital twin within a multiphysics simulation environment. The simulation geometry defines a segment of the blood vessel as a fluid domain, bounded by realistic endothelial tissue layers. Within this volume, discrete cellular models are populated according to physiological distributions. Crucially, the material properties of each cellular compartment are assigned directly from the requisite electromagnetic characterization library, ensuring that the simulation accurately reflects the frequency-dependent behavior of real biological tissue. The setup is finalized by embedding virtual transmitter (Tx) and receiver (Rx) antennas at opposite ends of the vessel segment.

Upon establishing the simulation environment, the system executes a series of spectral sweeps to emulate the ISAC sensing process. A broadband electromagnetic signal is excited at the transmitter port, prompting the solver to compute the propagation of these waves through the heterogeneous blood medium by explicitly solving Maxwell’s equations. The cumulative result of this physical interaction is captured at the receiver as the complex Channel Frequency Response (CFR), denoted as $H(f,L)$. This raw dataset encapsulates the frequency-dependent attenuation and cellular interaction phenomena, as mathematically formalized in Eq. (\ref{eq:freq_response}), thereby effectively encoding the biological state of the blood directly into the communication signal.

Following data acquisition, the extraction of biological features begins in the time domain by converting the CFR into the CIR. Since the Effective Medium Theory governs the channel, we analyze the Pulse Dispersion Characteristics. The primary metric is the Pulse Broadening Factor; as the concentration of formed elements increases, the effective permittivity of the medium rises, introducing stronger frequency-dependent group delay that widens the received pulse envelope. Additionally, the Pulse Amplitude Attenuation serves as a direct proxy for the bulk conductivity of the medium.

In the frequency domain, the analysis targets the spectroscopic signatures of the biological constituents. The Spectral Tilt provides a robust measure for distinguishing cell types; different cells exhibit unique dielectric relaxation frequencies, causing the signal to attenuate unevenly across the bandwidth. Furthermore, the Group Delay Variation is analyzed. This metric, derived from the derivative of the phase response, reveals subtle shifts in the refractive index of the surrounding plasma.

Crucially, the time and frequency domain metrics detailed above constitute a foundational rather than an exhaustive feature set. The complex, dispersive nature of the biological medium embeds a high density of information within the channel, offering numerous auxiliary metrics that can be exploited for deeper characterization. 

Finally, while the fundamental architecture assumes a single-antenna link, the framework is intrinsically extensible to Multi-Input Multi-Output (MIMO) configurations. This architectural evolution would leverage spatial diversity to significantly expand the dimensionality of the feature space, providing a richer, high-resolution dataset to train the AI models for even more granular biological classification.

\subsection{AI-Driven Biological State Classification}
The final stage of the framework employs Artificial Intelligence (AI) to map the extracted features to actionable hematological insights. Since obtaining large-scale in-vivo datasets for training is challenging, the multiphysics simulation environment functions as a digital twin, generating a comprehensive library of synthetic labeled data to train the models.

\subsubsection{CBC Parameter Detection} The main aim of this module is to measure the cellular components of blood that make up a routine Complete Blood Count (CBC). Changes in these values serve as key markers for disorders such as anemia (reduced RBCs/hemoglobin), leukemia or ongoing infection (abnormal WBC levels), and thrombocytopenia (decreased platelet count).

To obtain accurate estimates without relying on physical samples, the system employs a supervised learning framework based on regression techniques. During training, a multiphysics simulation environment serves as a digital twin. By systematically adjusting the concentrations of blood components in this simulated setting, a labeled dataset can be constructed. The inputs of this dataset are the extracted electromagnetic features, and the outputs are the corresponding known cell counts.

Because the dielectric characteristics of blood vary dynamically with cell density, the model can be trained to capture the complex, non-linear relationship between these signal perturbations and cellular concentrations. During inference, when a new unlabeled sample is acquired, the pre-trained model outputs continuous numerical estimates for each blood constituent. Since several cell populations are present at the same time and jointly shape the ISAC signal, multi-parameter regression methods can be used to disentangle their respective contributions and to simultaneously infer the complete CBC profile.

\subsubsection{Biomarker Detection} While CBC detection focuses on quantification, the Biomarker Detection module focuses on qualitative classification and anomaly identification. The foundational strategy involves establishing a robust baseline model representing normal physiological states. This can be achieved by training the algorithm on simulated data representing regular blood constituents within healthy reference ranges.

To enable the detection of specific pathologies, the simulation environment can then configured to introduce pathological biomarkers. These may include morphological changes, the presence of foreign pathogens, or abnormal plasma composition. The system extracts specific features that correlate with these anomalies.

The resulting dataset, containing both healthy and unhealthy-labeled scenarios, can be used to train a classifier. During continuous monitoring, the system compares real-time extracted features against this learned manifold. If the signal features diverge from the healthy baseline and align with the signature of a simulated biomarker, the system flags the specific disease. This classification task typically employs algorithms such as Support Vector Machines (SVM) for binary healthy/unhealthy determination, or Deep Neural Networks for multi-class categorization of specific conditions.

\subsection{Practical Implementation}
\label{sec:practical}

Although mathematical models and theoretical simulations provide a foundational framework for ISAC in continuous blood monitoring, the physical implementation of such systems hinges on parallel advancements in nanotechnology and material science. Here, we synthesize key developments in nanotechnology, materials science, and bioelectronics to bridge the gap between theory and practice. Specifically, translating these mathematical abstractions into deployable in-vivo systems necessitates addressing three fundamental hardware challenges: antenna miniaturization, biocompatibility, and energy harvesting.

\subsubsection{Antenna Design}
The physical implementation of an intravascular antenna is tightly constrained by the geometric limits of the human microvasculature. To avoid embolic complications and maintain unobstructed capillary perfusion, the largest permissible device dimension is on the order of 5–10 $\mu m$. This requirement poses a fundamental problem for conventional antenna theory, which stipulates that efficient radiation demands a physical size on the order of the operating wavelength. If a traditional metallic dipole were miniaturized to fit within this size envelope, its resonant frequency would shift into the optical regime (15–30 THz), where strong absorption by biological tissues prevents any meaningful communication \cite{llatser_2012_graphenebased}.

To resolve this challenge, antenna design has pivoted toward nanomaterials that support plasmonic modes and slow-wave propagation \cite{llatser_2012_graphenebased}. These materials confine electromagnetic fields to the antenna surface, allowing for resonance in the Terahertz (THz) band (0.1--10 THz) despite the microscopic physical dimensions of the radiator.

Among these materials, graphene has emerged as the quintessential material for intrabody nano-communication due to its ability to support Surface Plasmon Polaritons in the THz band. The propagation velocity of SPPs on graphene is significantly slower than the speed of light, effectively compressing the wavelength. Consequently, a graphene nano-ribbon with a length of only 1--5 $\mu m$ can support a standing wave corresponding to a free-space frequency of 1--10 THz \cite{ullah_2020_review}. This shifts the operating frequency from the highly absorptive optical range to the THz window, offering manageable path loss for short-range links.

A distinct advantage of graphene is its dynamic tunability. The material's optical conductivity is governed by its chemical potential, which can be adjusted via electrostatic gating. Research indicates that varying chemical potential allows the antenna to dynamically hop frequencies across a wide bandwidth \cite{dash_2020_nanoantennas}. Such reconfigurability is indispensable for ISAC frameworks, facilitating the evasion of high-loss water absorption bands while permitting the spectral sweeping needed for feature extraction.

\subsubsection{Biocompatibility}
Achieving biocompatibility presents a significant challenge for intravascular hardware. Upon entering the bloodstream, sensors typically elicit a foreign body response. To mitigate this reaction, two primary strategies have emerged: passive shielding via hydrophilic barriers \cite{wang_2025_bioinspired} and active camouflage through biomimicry \cite{zhao_2025_application}.

Passive protection is often achieved using zwitterionic polymers. These materials possess balanced cationic and anionic groups that exhibit strong hydrophilicity \cite{wang_2025_bioinspired}. This property allows the formation of a dense, tightly bound hydration shell around the device. By creating this stable water layer, the polymer prevents proteins and cells from displacing the water and adhering to the sensor surface, effectively masking the device from thrombotic and immune responses.

Conversely, cell membrane cloaking utilizes an active disguise strategy. This technique involves encapsulating the sensor within the membranes of natural cells \cite{zhao_2025_application}. For example, red blood cells (RBCs) circulate for extended periods due to the expression of CD47, a transmembrane protein acting as a self-marker. This protein signals the immune system to inhibit phagocytosis (the "don't eat me" signal). By coating the sensor in an RBC membrane, the device acquires this immunomodulatory capability, enabling prolonged circulation and data collection.

\subsubsection{Energy Harvesting}
Achieving energy autonomy is essential for intravascular nanodevices, as battery replacement is challenging and space is limited. To overcome this, Hybrid Energy Harvesting Systems (HEHS) combine Triboelectric Nanogenerators (TENGs) and Glucose Biofuel Cells (GBFCs) to utilize the blood's biomechanical and biochemical resources \cite{li_2020_hybrid}. TENGs harvest kinetic energy from blood flow through liquid-solid interactions, while GBFCs provide a continuous baseload power via the electrochemical oxidation of glucose.

When environmental harvesting is insufficient, Wireless Power Transfer (WPT) serves as an external energy source, typically implemented through inductive coupling or ultrasonic transduction \cite{ibrahim_2018_comprehensive}. Inductive coupling relies on magnetic fields generated between coils to transfer energy, though efficiency in this mode scales closely with coil size and quality factor. Alternatively, ultrasonic power transfer utilizes acoustic pressure waves which propagate through tissue with low attenuation. Because the speed of sound in tissue is significantly slower than the speed of light, acoustic wavelengths are much shorter, allowing for efficient resonance in sub-millimeter receivers suitable for deep-tissue implantation.

\section{Applications}
\label{sec:5}
The integration of ISAC-enabled nanonetworks for continuous blood monitoring can catalyze a paradigm shift in healthcare, extending the utility of diagnostics from discrete clinical snapshots to continuous, dynamic data streams. By digitizing biological signals and aggregating them within secure environments, this technology can unlock transformative applications across clinical, public, and societal domains.

\subsection{Continuous Inpatient Monitoring} In the acute care setting, the technology can serve as a blood monitor, revolutionizing inpatient protocols by replacing intermittent blood draws with real-time surveillance. Just as cardiac telemetry currently tracks electrical activity, this system could continuously evaluate the complete blood count (CBC) to track disease progression with granular precision. Such high-frequency data allows clinicians to monitor the trajectory of white blood cell counts minute-by-minute, enabling the immediate assessment of antibiotic efficacy in septic patients or the early detection of post-operative internal bleeding before clinical symptoms manifest. By providing a continuous stream of physiological data, the system can drastically reduce reaction times to critical events in intensive care units and post-surgical wards.

\subsection{Longitudinal Personalized Medicine} Beyond the hospital walls, the capability to archive and analyze historical data enables true personalized medicine through longitudinal health tracking. Rather than relying on population-wide averages, individuals can identify their unique baseline physiological markers, creating a personalized normal for their blood composition. This longitudinal dataset allows for the detection of subtle, sub-clinical deviations, such as gradual downward trends in hemoglobin indicating pre-anemia, or slight elevations in inflammatory markers signaling early immune responses. By visualizing these long-term patterns, the technology facilitates preventative interventions and lifestyle adjustments long before a condition escalates to the point of requiring hospitalization.

\subsection{Networked Epidemiology and Resource Optimization} By scaling this technology across a population, individual sensors can function as nodes in a distributed biological radar, creating a powerful tool for pandemic detection and management. If the specific electromagnetic signatures of bacterial or fungal pathogens can be identified, the aggregated data allows for the rapid triangulation of outbreak epicenters. This network could empower individuals through anonymized biological weather maps on their personal devices, allowing users to visualize high-risk zones of contagious disease and take proactive self-protection measures. Furthermore, this real-time epidemiological data can optimize the allocation of healthcare resources. Hospitals and pharmacies could utilize public health dashboards to predict surges in demand, ensuring that antibiotics, antivirals, and critical care beds are pre-positioned in regions showing early signals of rising infection rates.

\subsection{Psychosomatic and Societal Diagnostics} The capabilities of continuous monitoring can extend beyond physical pathology to the realm of mental health and societal well-being. By calibrating the system to detect hormonal fluctuations, it may be possible to infer the psychological state of the user. On an individual level, this could assist in the early detection and management of psychological disorders by correlating biological markers with mood patterns. On a macro scale, analyzing anonymized hormonal trends across a city or demographic could provide unprecedented insights into societal stress levels and collective mental health, enabling policymakers to assess the psychosomatic impact of economic or social events in real-time.

\section{Conclusion}
\label{sec:6}
The transition from discrete clinical snapshots to continuous, dynamic data streams represents a fundamental transformation in medical diagnostics. In this article, we presented a framework for ubiquitous hematological profiling driven by Integrated Sensing and Communication (ISAC), demonstrating how communication signals can be exploited to monitor blood in real-time, effectively converting them into diagnostic tools. The technical realizability of this paradigm relies on the rigorous electromagnetic characterization of biological cells to establish a ground truth for channel modeling. Furthermore, practical deployment hinges on parallel advancements in nanotech-enabled antenna design, biocompatibility, and energy harvesting.

Beyond technical feasibility, the integration of ISAC-enabled nanonetworks signals a paradigm shift in proactive healthcare. The possible applications are far-reaching. By identifying physiological irregularities at the very onset of cellular changes, this technology has the potential to revolutionize early diagnosis. In the long run, as these nanodevices become widely deployed within the Internet of Bio-Nano Things (IoBNT), they will not only protect individual well-being but also offer a large-scale perspective on public health, converting biological signals into practical, data-driven insights for society.

\bibliographystyle{IEEEtran}
\bibliography{references.bib}

\vspace{-5cm}
\begin{IEEEbiography}
    [{\includegraphics[width=1in, height=1.25in, clip, keepaspectratio]{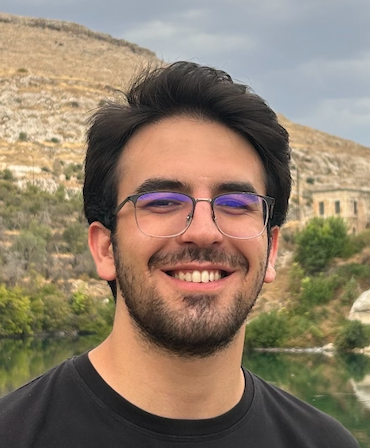}}]{Fatih Efe Bilgen} is a Ph.D. candidate in Engineering at the University of Cambridge, where he works under the supervision of Prof. Akan. He completed his B.Sc. in Electrical and Electronics Engineering with a dual degree in Mathematics from Koç University, Istanbul, Turkey, in June 2024, followed by an M.S. degree in Electrical and Electronics Engineering in July 2025 from the same institution. His research focuses on advancing the Internet of Everything through innovative approaches.
\end{IEEEbiography}

\vspace{-7cm}
\begin{IEEEbiography}
[{\includegraphics[width=1in,height=1.25in,clip,keepaspectratio]{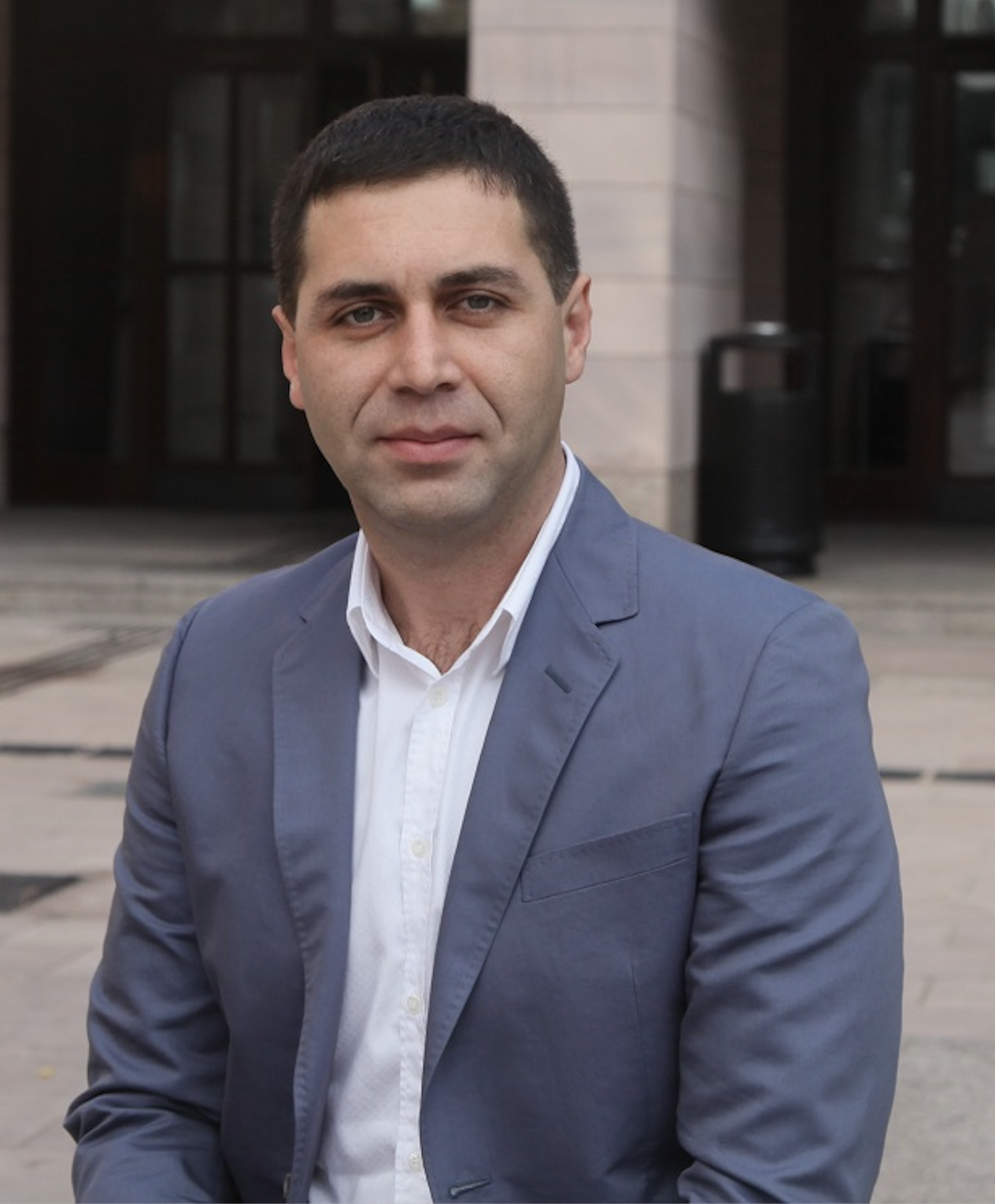}}]{Ozgur B. Akan (Fellow, IEEE)}
received the PhD from the School of Electrical and Computer Engineering Georgia Institute of Technology Atlanta, in 2004. He is currently the Head of Internet of Everything (IoE) Group, with the Department of Engineering, University of Cambridge, UK and the Director of Centre for neXt-generation Communications (CXC), Koç University, Turkey. His research interests include wireless, nano, and molecular communications and Internet of Everything.
\end{IEEEbiography}

\end{document}